\documentclass[twocolumn]{aastex63}

\newcommand{\hei}{He\,I}

\usepackage[utf8]{inputenc} 
\usepackage[T1]{fontenc}
\newcommand{\unit}[1]{\ensuremath{\, \mathrm{#1}}}

\submitjournal{AJ}
\accepted{\today}

\shorttitle{High-resolution transit observations of TRAPPIST-1 planets}
\shortauthors{Krishnamurthy et al.}

\graphicspath{{./}{figures/}}

\begin{document}

\title{Non-detection of Helium in the upper atmospheres of TRAPPIST-1b, e and f 
\footnote{Based on data collected at Subaru Telescope, operated by the National Astronomical Observatory of Japan, Hobby–Eberly Telescope operated by The University of Texas McDonald Observatory, and ARC 3.5m Telescope at Apache Point Observatory.}}

\correspondingauthor{Vigneshwaran Krishnamurthy}
\email{krishnamurthy.v.aa@m.titech.ac.jp}

\author[0000-0003-2310-9415]{Vigneshwaran Krishnamurthy}
\affiliation{Department of Earth and Planetary Sciences, Tokyo Institute of Technology, 2-12-1 Ookayama, Meguro-ku, Tokyo 152-8551, Japan}

\author[0000-0003-3618-7535]{Teruyuki Hirano}
\affiliation{Astrobiology Center, NINS, 2-21-1 Osawa, Mitaka, Tokyo 181-8588, Japan}
\affiliation{National Astronomical Observatory of Japan, NINS, 2-21-1 Osawa, Mitaka, Tokyo 181-8588, Japan}

\author[0000-0001-7409-5688]{Guðmundur Stefánsson}
\affiliation{Department of Astrophysical Sciences, Princeton University, 4 Ivy Lane, Princeton, NJ 08540, USA}
\affiliation{Henry Norris Russell Fellow}

\author[0000-0001-8720-5612]{Joe P. Ninan}
\affiliation{Department of Astronomy \& Astrophysics, 525 Davey Laboratory, Pennsylvania State University, University
Park, PA, 16802, USA}
\affiliation{Center for Exoplanets and Habitable Worlds, 525 Davey Laboratory, Pennsylvania State University,University
Park, PA, 16802, USA}

\author[0000-0001-9596-7983]{Suvrath Mahadevan}
\affiliation{Department of Astronomy \& Astrophysics, 525 Davey Laboratory, Pennsylvania State University, University
Park, PA, 16802, USA}
\affiliation{Center for Exoplanets and Habitable Worlds, 525 Davey Laboratory, Pennsylvania State University,University
Park, PA, 16802, USA}

\author[0000-0002-5258-6846]{Eric Gaidos}
\affiliation{Department of Earth Sciences, University of Hawai'i at M\={a}noa, Honolulu, HI 96822, USA}
\affiliation{Center for Space and Habitability, University of Bern, Gesellschaftsstrasse 6, 3012 Bern, Switzerland}

\author[0000-0002-5893-2471]{Ravi Kopparapu}
\affiliation{NASA Goddard Space Flight Center, 8800 Greenbelt Road, Greenbelt, MD 20771, USA}

\author{Bunei Sato}
\affiliation{Department of Earth and Planetary Sciences, Tokyo Institute of Technology, 2-12-1 Ookayama, Meguro-ku, Tokyo 152-8551, Japan}

\author[0000-0003-4676-0251]{Yasunori Hori}
\affiliation{Astrobiology Center, NINS, 2-21-1 Osawa, Mitaka, Tokyo 181-8588, Japan}
\affiliation{National Astronomical Observatory of Japan, NINS, 2-21-1 Osawa, Mitaka, Tokyo 181-8588, Japan}

\author[0000-0003-4384-7220]{Chad F. Bender}
\affiliation{Steward Observatory, The University of Arizona, 933 N. Cherry Ave, Tucson, AZ 85721, USA}

\author[0000-0003-4835-0619]{Caleb I. Ca\~nas}
\affiliation{NASA Earth and Space Science Fellow}
\affiliation{Department of Astronomy \& Astrophysics, 525 Davey Laboratory, Pennsylvania State University, University
Park, PA, 16802, USA}
\affiliation{Center for Exoplanets and Habitable Worlds, 525 Davey Laboratory, Pennsylvania State University,University
Park, PA, 16802, USA}

\author[0000-0002-2144-0764]{Scott A. Diddams}
\affil{Time and Frequency Division, National Institute of Standards and Technology, 325 Broadway, Boulder, CO 80305, USA}
\affil{Department of Physics, University of Colorado, 2000 Colorado Avenue, Boulder, CO 80309, USA}

\author[0000-0003-1312-9391]{Samuel Halverson}
\affil{Jet Propulsion Laboratory, 4800 Oak Grove Drive, Pasadena, CA 91109, USA}

\author[0000-0002-7972-0216]{Hiroki Harakawa}
\affiliation{Subaru Telescope, 650 N. Aohoku Place, Hilo, HI 96720, USA}

\author[0000-0002-6629-4182]{Suzanne Hawley}
\affil{Department of Astronomy, Box 351580, University of Washington, Seattle, WA 98195, USA}

\author[0000-0002-1664-3102]{Fred Hearty}
\affiliation{Department of Astronomy \& Astrophysics, 525 Davey Laboratory, Pennsylvania State University, University
Park, PA, 16802, USA}
\affiliation{Center for Exoplanets and Habitable Worlds, 525 Davey Laboratory, Pennsylvania State University,University
Park, PA, 16802, USA}

\author[0000-0003-1263-8637]{Leslie Hebb}
\affiliation{Department of Physics, Hobart and William Smith Colleges, 300 Pulteney Street, Geneva, NY, 14456, USA}

\author[0000-0003-0786-2140]{Klaus Hodapp}
\affiliation{University of Hawaii, Institute for Astronomy, 640 N. Aohoku Place, Hilo, HI 96720, USA}

\author{Shane Jacobson}
\affiliation{University of Hawaii, Institute for Astronomy, 640 N. Aohoku Place, Hilo, HI 96720, USA}

\author[0000-0001-8401-4300]{Shubham Kanodia}
\affiliation{Department of Astronomy \& Astrophysics, 525 Davey Laboratory, Pennsylvania State University, University
Park, PA, 16802, USA}
\affiliation{Center for Exoplanets and Habitable Worlds, 525 Davey Laboratory, Pennsylvania State University,University
Park, PA, 16802, USA}

\author[0000-0003-0114-0542]{Mihoko Konishi}
\affiliation{Faculty of Science and Technology, Oita University, 700 Dannoharu, Oita 870-1192, Japan}

\author[0000-0001-6181-3142]{Takayuki Kotani}
\affiliation{Astrobiology Center, NINS, 2-21-1 Osawa, Mitaka, Tokyo 181-8588, Japan}
\affiliation{National Astronomical Observatory of Japan, NINS, 2-21-1 Osawa, Mitaka, Tokyo 181-8588, Japan}
\affiliation{Department of Astronomy, School of Science, The Graduate University for Advanced Studies (SOKENDAI), 2-21-1 Osawa, Mitaka, Tokyo, Japan}

\author[0000-0001-7458-1176]{Adam Kowalski}
\affil{Department of Astrophysical and Planetary Sciences, University of Colorado Boulder, 2000 Colorado Avenue, Boulder, CO 80305, USA}
\affil{National Solar Observatory, University of Colorado Boulder, 3665 Discovery Drive, Boulder, CO 80303, USA}
\affil{Laboratory for Atmospheric and Space Physics, University of Colorado Boulder, 3665 Discovery Drive, Boulder, CO 80303, USA}

\author[0000-0002-9294-1793]{Tomoyuki Kudo}
\affiliation{Subaru Telescope, 650 N. Aohoku Place, Hilo, HI 96720, USA}

\author{Takashi Kurokawa}
\affiliation{Astrobiology Center, NINS, 2-21-1 Osawa, Mitaka, Tokyo 181-8588, Japan}
\affiliation{Institute of Engineering, Tokyo University of Agriculture and Technology, 2-24-16, Nakacho, Koganei, Tokyo, 184-8588, Japan}

\author[0000-0002-4677-9182]{Masayuki Kuzuhara}
\affiliation{Astrobiology Center, NINS, 2-21-1 Osawa, Mitaka, Tokyo 181-8588, Japan}
\affiliation{National Astronomical Observatory of Japan, NINS, 2-21-1 Osawa, Mitaka, Tokyo 181-8588, Japan}

\author[0000-0002-9082-6337]{Andrea Lin}
\affiliation{Department of Astronomy \& Astrophysics, 525 Davey Laboratory, Pennsylvania State University, University
Park, PA, 16802, USA}
\affiliation{Center for Exoplanets and Habitable Worlds, 525 Davey Laboratory, Pennsylvania State University,University
Park, PA, 16802, USA}

\author[0000-0001-8222-9586]{Marissa Maney}
\affiliation{Department of Astronomy \& Astrophysics, 525 Davey Laboratory, Pennsylvania State University, University
Park, PA, 16802, USA}

\author[0000-0001-5000-1018]{Andrew J. Metcalf}
\affiliation{Space Vehicles Directorate, Air Force Research Laboratory, 3550 Aberdeen Ave. SE, Kirtland AFB, NM 87117, USA}
\affiliation{Time and Frequency Division, National Institute of Standards and Technology, 325 Broadway, Boulder, CO 80305, USA} 
\affiliation{Department of Physics, University of Colorado, 2000 Colorado Avenue, Boulder, CO 80309, USA}

\author[0000-0003-2528-3409]{Brett Morris}
\affiliation{Center for Space and Habitability, University of Bern, Gesellschaftsstrasse 6, 3012 Bern, Switzerland}

\author[0000-0001-9326-8134]{Jun Nishikawa}
\affiliation{National Astronomical Observatory of Japan, NINS, 2-21-1 Osawa, Mitaka, Tokyo 181-8588, Japan}
\affiliation{Department of Astronomy, School of Science, The Graduate University for Advanced Studies (SOKENDAI), 2-21-1 Osawa, Mitaka, Tokyo, Japan}
\affiliation{Astrobiology Center, NINS, 2-21-1 Osawa, Mitaka, Tokyo 181-8588, Japan}

\author{Masashi Omiya}
\affiliation{Astrobiology Center, NINS, 2-21-1 Osawa, Mitaka, Tokyo 181-8588, Japan}
\affiliation{National Astronomical Observatory of Japan, NINS, 2-21-1 Osawa, Mitaka, Tokyo 181-8588, Japan}

\author[0000-0003-0149-9678]{Paul Robertson}
\affil{Department of Physics and Astronomy, The University of California, Irvine, Irvine, CA 92697, USA}

\author[0000-0001-8127-5775]{Arpita Roy}
\affil{Space Telescope Science Institute, 3700 San Martin Dr., Baltimore, MD 21218, USA}
\affil{Department of Physics and Astronomy, Johns Hopkins University, 3400 N Charles St, Baltimore, MD 21218, USA}

\author[0000-0002-4046-987X]{Christian Schwab}
\affil{Department of Physics and Astronomy, Macquarie University, Balaclava Road, North Ryde, NSW 2109, Australia}

\author{Takuma Serizawa}
\affiliation{Institute of Engineering, Tokyo University of Agriculture and Technology, 2-24-16, Nakacho, Koganei, Tokyo, 184-8588, Japan}

\author[0000-0002-6510-0681]{Motohide Tamura}
\affiliation{Astrobiology Center, NINS, 2-21-1 Osawa, Mitaka, Tokyo 181-8588, Japan}
\affiliation{National Astronomical Observatory of Japan, NINS, 2-21-1 Osawa, Mitaka, Tokyo 181-8588, Japan}
\affiliation{Department of Astronomy, Graduate School of Science, The University of Tokyo, 7-3-1 Hongo, Bunkyo-ku, Tokyo 113-0033, Japan}

\author{Akitoshi Ueda}
\affiliation{National Astronomical Observatory of Japan, NINS, 2-21-1 Osawa, Mitaka, Tokyo 181-8588, Japan}

\author[0000-0003-4018-2569]{S\'ebastien Vievard}
\affiliation{Astrobiology Center, NINS, 2-21-1 Osawa, Mitaka, Tokyo 181-8588, Japan}
\affiliation{Subaru Telescope, 650 N. Aohoku Place, Hilo, HI 96720, USA}

\author[0000-0001-9209-1808]{John Wisniewski}
\affiliation{Homer L. Dodge Department of Physics and Astronomy, University of Oklahoma, 440 W. Brooks Street, Norman, OK 73019, USA}

\begin{abstract}
We obtained high-resolution spectra of the ultra-cool M-dwarf TRAPPIST-1 during the transit of its planet `b' using two high dispersion near-infrared spectrographs, IRD instrument on the Subaru 8.2m telescope and HPF instrument on the 10m Hobby-Eberly Telescope. 
These spectroscopic observations are complemented by a photometric transit observation for planet `b' using the APO/ARCTIC, which assisted us to capture the correct transit times for our transit spectroscopy. 
Using the data obtained by the new IRD and HPF observations, as well as the prior transit observations of planets `b', `e' and `f' from IRD, we attempt to constrain the atmospheric escape of the planet using the He\,I triplet 10830 \AA{} absorption line. We do not detect
evidence for any primordial extended H-He atmospheres in all three planets. 
To limit any planet related absorption, we place an upper limit on the equivalent widths of <7.754 m\AA{} for planet `b', <10.458 m\AA{} for planet `e', and <4.143 m\AA{} for planet `f' at 95\% confidence from the IRD data, and <3.467 m\AA{} for planet `b' at 95\% confidence from HPF data. Using these limits along with a solar-like composition isothermal Parker wind model, we attempt to constrain the mass-loss rates for the three planets. For TRAPPIST-1b, our models exclude the highest possible energy-limited rate for a wind temperature $<$5000 K. This non-detection of extended atmospheres having low mean-molecular weight in all three planets aids in further constraining their atmospheric composition by steering the focus towards the search of high molecular weight species in their atmospheres.

\end{abstract}

\keywords{High resolution spectroscopy (2096) --- 
Exoplanet evolution (491)}

\section{Introduction} \label{sec:intro}
In the last two decades, the exoplanet research has ventured into the field of atmospheric characterization, exploring and understanding the planet's atmosphere through transmission and emission spectroscopy. Lyman-$\alpha$ in the UV has been the primary tracer for hydrogen-rich atmospheres, e.g., HD 209458b \citep{2003Natur.422..143V}, GJ 436b \citep{2014ApJ...786..132K, 2015Natur.522..459E}, GJ 3470b \citep{2018A&A...620A.147B}. 
 
In addition to Ly-$\alpha$, the Helium I triplet at 10830 \AA{} has been proposed to be an ideal marker at near-infrared (NIR) wavelengths \citep{2000ApJ...537..916S}. Unlike Ly-$\alpha$, this He\,I triplet can be observed from a ground-based facility using a high-resolution NIR spectrograph. The recent theoretical study by \cite{2018ApJ...855L..11O} and, more specifically, for K-dwarfs and M-dwarfs in \cite{2019ApJ...881..133O} have paved the way for better constraints on the atmospheric escape rates. Hubble Space Telescope's Wide Field Camera 3 (\emph{HST/WFC3}) first detected this transmission signal in WASP-107b \citep{2018Natur.557...68S}, which was independently verified through high resolution ground-based observations \citep{2019A&A...623A..58A, 2020AJ....159..115K}, estimating an absorption level of $5.54 \pm 0.27\,\%$. Following this, Helium in several planets' extended atmospheres have been detected, namely, HAT-P-11b \citep{2018Sci...362.1384A, 2018ApJ...868L..34M}, HD 189733b \citep{2018A&A...620A..97S, 2020A&A...639A..49G}, WASP-69b \citep{2018Sci...362.1388N}, HD 209458b \citep{2019A&A...629A.110A}, GJ 3470b \citep{2020ApJ...894...97N, 2020A&A...638A..61P} and HAT-P-18b \citep{2021ApJ...909L..10P}. In addition, non-detection upper limits of He\,I absorption have been reported for GJ 436b \& Kelt-9b \citep{2018Sci...362.1388N}, WASP-12b \citep{2018RNAAS...2...44K}, GJ 1214b \citep{2019RNAAS...3...24C}, K2-100b \citep{2020MNRAS.495..650G}, AU Mic b \citep{2020ApJ...899L..13H} and K2-25b \citep{2020MNRAS.498L.119G} through transmission spectra. Detection through emission spectra was attempted in $\tau$ Boo b, but He\,I was not detected \citep{2020A&A...641A.161Z}. \cite{2020AJ....159..278V} used a novel ultra-narrowband diffuser-assisted photometric technique to estimate the transit depth at the \hei\, and presented an upper limit in the atmosphere of WASP-52b.
 
 The stellar radiation plays a crucial role in the Helium absorption levels in these planets. The He\,I lines originates from the metastable $2^3$S triplet state, where the atoms are populated by the process of recombination of ionized Helium. Photons blueward of 50.4 nm ionize the Helium atoms, therefore planets receiving enhanced X-ray and extreme-ultraviolet (EUV) flux are expected to produce an amplified Helium absorption \citep{2018ApJ...855L..11O}. In addition to collisional de-excitation, the mid-ultraviolet (mid-UV) photons around 260 nm can ionize triplet state $2^3S$ atoms, thereby de-populating the energy level. Therefore, the ratio of EUV to mid-UV governs the atoms available for this absorption and the strength of the absorption signal \citep{2019ApJ...881..133O}. K-dwarfs, with stronger EUV flux and lower mid-UV flux are the ideal targets for this transit absorption signal at 1083\,nm. Hitherto, the exoplanets with detected He\,I triplet absorption confirm this trend, with five out of seven orbiting K-type stars. In the case of M-dwarfs, although having the hardest spectra for triplet Helium production at a fixed orbital distance, the EUV flux is at least an order of magnitude lower than that of K-type stars. This indicates that the M-dwarf planets need to be much more closer to their host star to produce the same level of triplet Helium population \citep{2019ApJ...881..133O}. More observations can help us better constrain the factors affecting this He\,I absorption level \citep{2020A&A...641A.161Z}.

 Since the discovery of the seven planets orbiting the ultra-cool dwarf TRAPPIST-1 \citep{2016Natur.533..221G, 2017Natur.542..456G}, the system has caught the attention of the exoplanet community. The planets of this old system \citep[$7.6\pm2.2$ Gyr;][]{2017ApJ...845..110B} are in a complex chain of resonant orbits with three planets in the star's habitable zone \citep{2017NatAs...1E.129L}. Its low mass \citep[$\sim0.09M_{\sun}$;][]{2018ApJ...853...30V} 
 and high multiplicity of seven Earth-sized transiting planets \citep{2018A&A...613A..68G} made it the primary target for several observational programs. Several transits of TRAPPIST-1 planets have been observed with the \textit{Spitzer Space Telescope} \citep{2018MNRAS.475.3577D, 2020A&A...640A.112D, 2020arXiv201001074A}, \emph{HST} \citep{2016Natur.537...69D, 2017A&A...599L...3B, 2018NatAs...2..214D, 2019AJ....157...11W}, K2-mission \citep{2017NatAs...1E.129L} and ground-based observatories \citep{2018NatAs...2..344G, 2020ApJ...890L..27H}. From \textit{HST}/WFC3 observations, \cite{2016Natur.537...69D} and \cite{2018NatAs...2..214D} claim that the planets TRAPPIST-1b, c, d, e and f lack cloud-free hydrogen-dominated atmospheres. \cite{2017A&A...599L...3B} detected a marginal flux decrease during their Ly-$\alpha$ measurements of TRAPPIST-1b transit using \textit{HST}'s Space Telescope Imaging Spectrograph (STIS), but could not confirm the presence of an extended atmosphere. \cite{2020ApJ...889...77H} found that all the primordially accreted hydrogen-rich atmospheres of TRAPPIST-1 planets would have been lost via hydrodynamic escape in the first few hundred Myr of planet formation.
 
 Here we present transit measurements of TRAPPIST-1b planet using Subaru Telescope's Infrared Doppler (IRD) instrument and Hobby-Eberly Telescope's (HET's) Habitable Zone Planet Finder (HPF) instrument. In addition to the new data obtained in this work, we also analyze the past IRD data obtained by \cite{2020ApJ...890L..27H}. We present the details of our observation and reduction strategy in Section 2. In Section 3, we discuss our key results, followed by discussion and conclusion in Section 4.
 
\section{Observations and data reduction} \label{sec2}
 
\subsection{Subaru/IRD} \label{sec2sub1}
We observed the transit of TRAPPIST-1b on the night of UT 2020 September 17 using IRD, which has a spectral resolution of $\approx 70,000$ in the operating wavelength range of $0.95-1.75$ $\mu m$, mounted on the 8.2-m Subaru Telescope on Maunakea, Hawaii \citep{2012SPIE.8446E..1TT, 2018SPIE10702E..11K}. This observation was carried out as a part of 2 open-use observation programs: S20B-069 (PI: Krishnamurthy) and S20A-UH104 (PI: Gaidos). The exposure time was set to 300 seconds. We obtained 21 frames of TRAPPIST-1.  We ended the observation with 3 frames of a rapidly-rotating A1 star, HIP 117774 \citep{1999mctd.book.....H} to aid in telluric corrections. On that particular night, we had high winds and moderate seeing of $\approx 0.5\farcs$ Wavelengths were simultaneously calibrated by injecting the light from Laser-Frequency Comb (LFC) into the spectrograph using a second fiber. 
 
The raw IRD data were reduced using the {\tt IRAF} software, followed by the precise wavelength calibration based on the LFC spectra using our own custom codes \citep[see][]{2020PASJ..tmp..243H}. On the transit night, the extracted 1D spectra had a per-pixel signal-to-noise ratio (S/N) of 19-20 at 1.0 $\unit{\mu m}$.
 
We also used the data from \cite{2020ApJ...890L..27H} in our analysis, where the transit of TRAPPIST-1b, e and f were captured. The primary focus of their paper was the Rossiter–McLaughlin (RM) effect measurements and not atmosphere characterization. The per-pixel S/N ratios for both these observations were similar ($\sim 19-20$ at 1 $\unit{\mu m}$) and were from the same instrument, IRD.

\subsection{HET/HPF} \label{sec2sub2}
HPF is a fiber-fed high resolution ($R=55,000$) spectrograph \citep{mahadevan2012,mahadevan2014} on the 10-m HET at McDonald Observatory in Texas. HPF is actively temperature and pressure stabilized to ensure a stable line profile and enable high precision radial velocities in the NIR \citep{stefansson2016}. All observations were carried out as part of the HET queue \citep{shetrone2007}. HPF also has a NIR LFC calibrator to provide a precise wavelength solution and track instrumental drifts at the $\sim$20 cm s$^{-1}$ 
level in 10-min bins \citep{metcalf2019}. Following \cite{stefansson2020} and \cite{stefansson2020b}, we elected not to use the simultaneous LFC calibration for any of the observations to minimize risk of contaminating the science spectrum from scattered light from the LFC. Instead, all drift corrections followed the methodology presented in \cite{stefansson2020}, where LFC frames that are taken periodically throughout the night are applied to derive a precise wavelength solution at the 30 cm s$^{-1}$ level. The HPF 1D spectra were processed using the HPF pipeline, using the algorithms presented in \cite{ninan2018}, \cite{kaplan2018}, and \cite{metcalf2019}. Barycentric corrections were calculated using \texttt{barycorrpy} \citep{kanodia2018}, following the algorithm from \citep{2014PASP..126..838W}.

Using HPF, we obtained three transits on the nights of UT 2018 October 2, 2019 July 31, and 2020 September 5. We obtained six 315-second exposures during transits for the first and third nights, with only three exposures for the second night due to bad weather. These spectra had a median S/N ratio of 38, 14, and 31 at $1.0\unit{\mu m}$ per 1D extracted pixel, respectively. Due to low S/N of the data for the second transit, we did not use those spectra for the subsequent analysis. In addition to the in-transit data, we obtained out-of-transit observations from separate nights to create a high S/N out-of-transit reference spectrum from 68 out-of-transit spectra obtained with an exposure time of 315 sec and 13 other spectra with an exposure time between 415 sec and 1224 sec. The reference spectrum had a total S/N of 271.

\subsection{APO/ARCTIC} \label{sec2sub3}
We observed a photometric transit of TRAPPIST-1b on the night of UT 2020 September 5 using the ARCTIC camera on the 3.5m Astrophysical Research Consortium (ARC) Telescope at Apache Point Observatory. We adopted the SDSS $i^\prime$ filter using an exposure time of 18 sec in the $4 \times 4$ binning fast-readout mode. In this mode, ARCTIC has a gain of $2 \unit{e^-/ADU}$ and a plate scale of $0\farcs44$ per pixel. 
We defocused the telescope to yield a target star FWHM of 9 pixels or $\sim4\farcs2$. To extract the photometry we used \texttt{AstroImageJ} following the methodology in \cite{stefansson2017}. We cleaned the data of cosmic rays using the \texttt{astroscrappy} code \citep{astroscrappy,vanDokkum2001} following \cite{stefansson2017}. We used an aperture radius of 10 pixels, and a sky-annulus with inner and outer radii of 20 pixels and 50 pixels, respectively.

\section{Analysis and Results} \label{sec3}
 
\subsection{Transit ephemerides} \label{sec3sub1}

\begin{figure}
\centering
\includegraphics[width=8.6 cm]{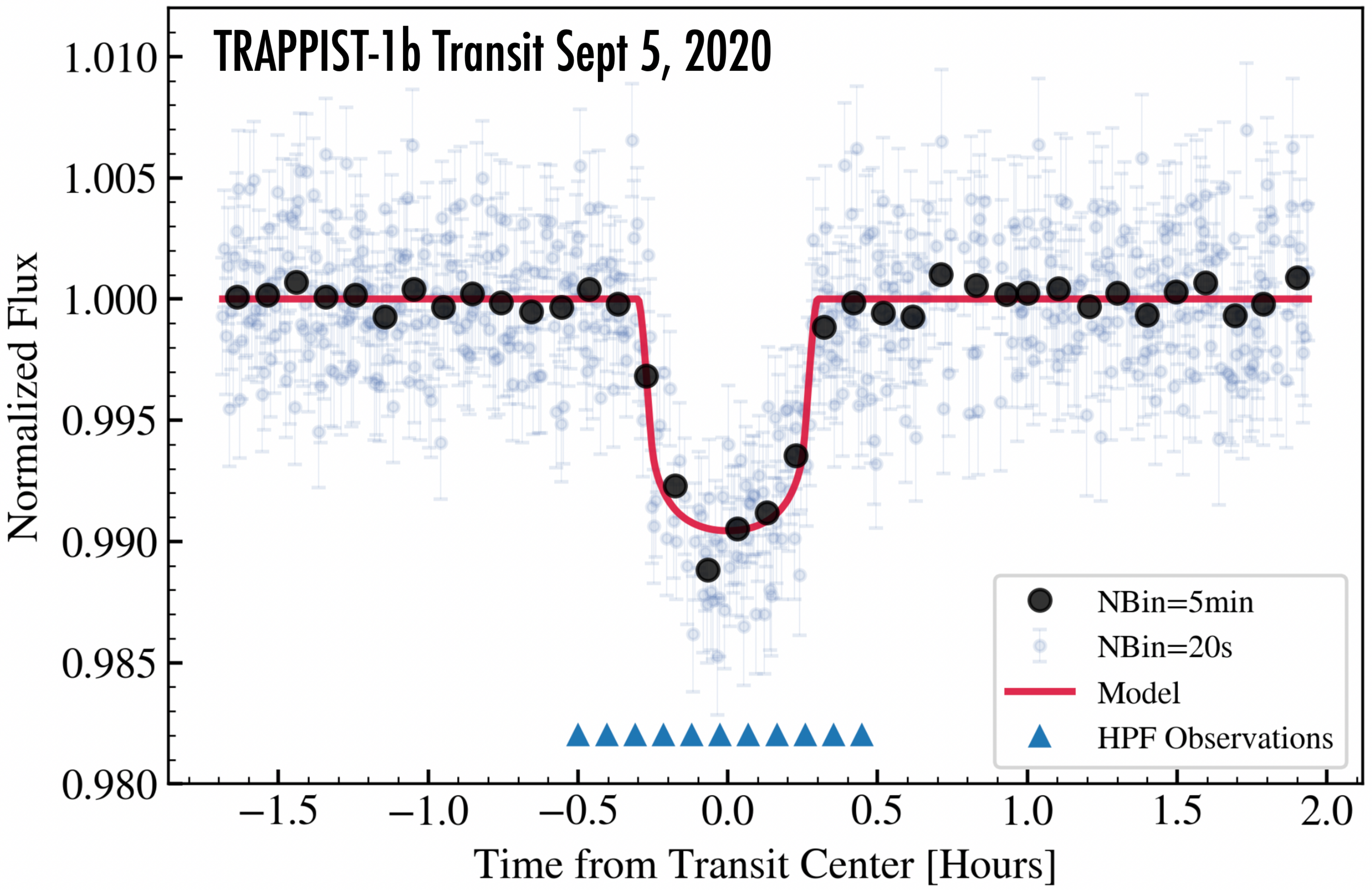}
\caption{Transit of TRAPPIST-1b observed with the ARCTIC instrument (black and blue points), and HPF (blue triangles) on the night of September 5, 2020.}
\label{fig:transit}
\end{figure}

We used the photometric data with ARCTIC/APO (Figure \ref{fig:transit}) to derive a precise transit midpoint for the 2020 September 5 transit observations with HPF/HET. To extract the transit midpoint, we fit the data using the \texttt{juliet} package \citep{Espinoza2019}, broadly following the methodology in \cite{stefansson2020b}. In short, we placed uninformative priors on the transit midpoint and transit depth, while placing Gaussian priors on the known orbital period, $a/R_*$, impact parameter from \cite{2017Natur.542..456G}. We fixed the limb-darkening parameters to the expected values in the SDSS $i^\prime$ filter: $q_1=0.42$, and $q_2=0.23$. To account for correlated noise, we follow \cite{stefansson2020b} and use the \texttt{juliet} Approximate Matern-3/2 kernel implemented in the \texttt{celerite} package \citep{Foreman-Mackey2017}. We confirmed that a model without a Gaussian Process noise model results in a fully consistent transit midpoint. The photometric data shows no evidence of flares during the transit.

We estimate the mid-transit time to be $\mathrm{T_c}$ $(\mathrm{BJD_{TBD}})$ = $2459097.80328_{-0.00021}^{+0.00022}$ and period as $1.510879_{-0.00017}^{+0.00022}$ days. The mid-transit time differ from the recent TTV derived ephemeris \citep{2020arXiv201001074A} by less than 10 seconds. For the 2018 transit of TRAPPIST-1b, e and f from  \cite{2020ApJ...890L..27H}, we used the ephemerides mentioned in their paper.

\subsection{He\,I triplet analysis} \label{sec3sub2}
Each spectrum is transferred to the stellar rest frame by correcting for the barycentric velocity and systematic radial velocity (RV). Individual IRD and HPF spectra were adjusted to account for the Doppler shift variation of the planet, thereby shifting them to the planet's rest frame. During the transit of planet `b', the master in-transit spectrum was obtained by median-combining the 6+6 (from 2 transits; 2018 and 2020) individual transit spectra from IRD and HPF separately. We did not use the 2019 HPF transit data, as it was affected by bad weather. For producing the master out-of-transit IRD spectrum, we excluded the first 7 frames after the transit. This was done to eliminate the possibility of contaminating the out-of-transit spectra from a possible evaporating tail from the planet. We used all the available out-of-transit frames to create the median-combined HPF master out-of-transit frame.

\begin{figure*}
\centering
\includegraphics[width=18cm]{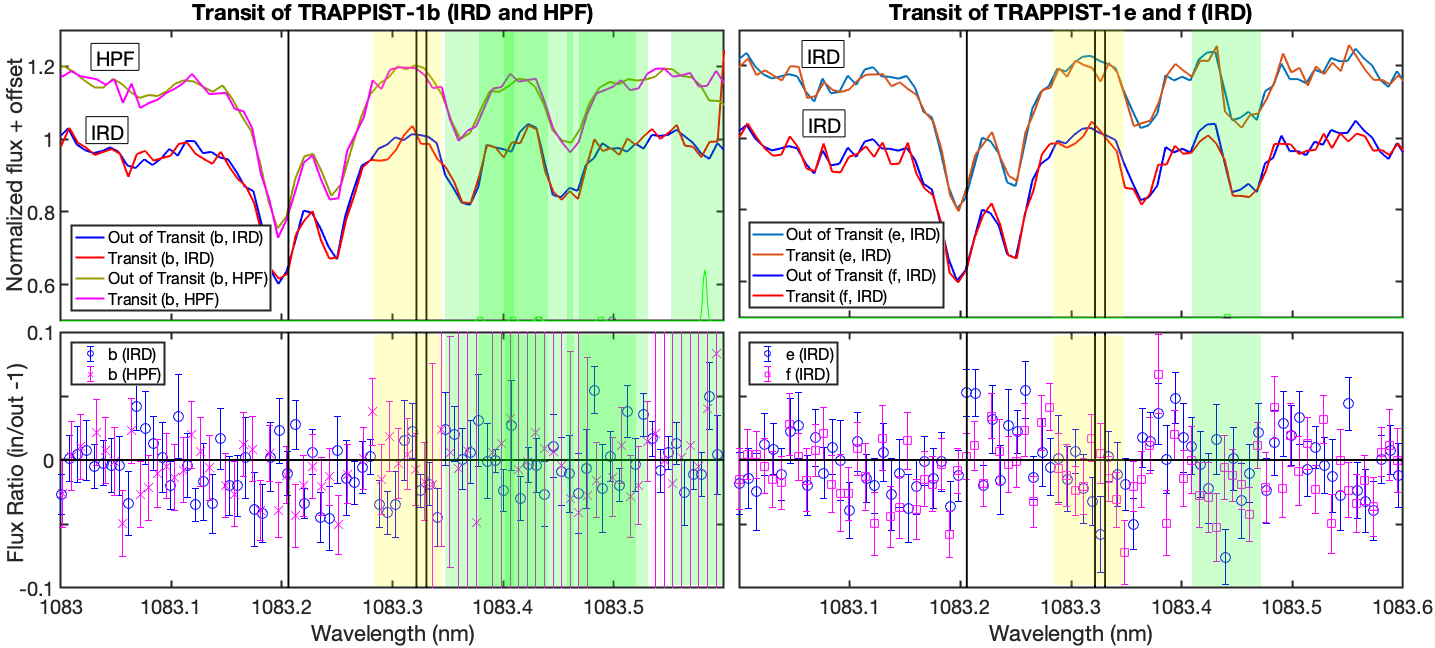}
\caption{Transmission spectra of TRAPPIST-1 planets. 
The top panels show the spectra of TRAPPIST-1 in the vicinity of He\,I triplet lines (solid black vertical lines). \textbf{Top left}: The in- and out-of-transit spectra of planet `b' from both IRD (red and blue) and HPF (magenta and light green) are shown. \textbf{Top right}: In- and out-of-transit spectra during the transit of planet `e' (orange and light blue) and during the transit of planet `f' (red and blue) from IRD \citep{2020ApJ...890L..27H}. \textbf{Bottom}: Their respective flux ratios (in/out -1). The green-shaded region corresponds to the OH-emission line position computed from \cite{2013A&A...560A..91J}. The computed known telluric $\unit{H_2O}$ positions from \cite{1973SoPh...28...15B} do not fall in the vicinity of Helium lines. The yellow-shaded region is our ``\textit{region of interest}''.}
\label{fig:fluxratio}
\end{figure*}

In the case of planets `e' and `f' transits captured by IRD in 2018, there could be possible overlap of the tail from planet `e' transit frames into the planet `f' transit frames, as the transit of planet `f' followed the transit of planet `e' immediately. The median combined master in-transit and out-of-transit spectra for planet `b' (from IRD and HPF), `e' (from IRD) and `f' (from IRD) are shown in the Figure \ref{fig:fluxratio}.

\subsubsection{Transmission spectra}
The stellar He\,I lines are not detected in IRD and HPF spectra (see top panels of Figure \ref{fig:fluxratio}). The absence of stellar lines is likely due to the low effective temperature of TRAPPIST-1 and not due to a line fill-in from collisional excitation \citep{2019A&A...632A..24F}.  From the IRD spectra of planet `b', we see a $\sim0.02$ nm wide feature about 0.03 nm blueward of the doublet center. This feature could originate from gaseous absorption from a Doppler-shifted tail moving relative to the star at $\sim-7$ km s$^{-1}$. To confirm that the feature originated during transit, we compared the strength of in-transit spectra with the same number of randomly sampled out-of-transit spectra. We performed this comparison for 1000 iterations. We find that the mean-strength of feature during transit (EW=5.17$\pm$3.18 m\AA{}) is higher than that of the random sampled out-of-transit spectra (EW=0.11$\pm$3.79 m\AA{}) in the IRD data. We do not detect such a blueward feature in the HPF spectra.

\begin{figure}
\centering
\includegraphics[width=8.6 cm]{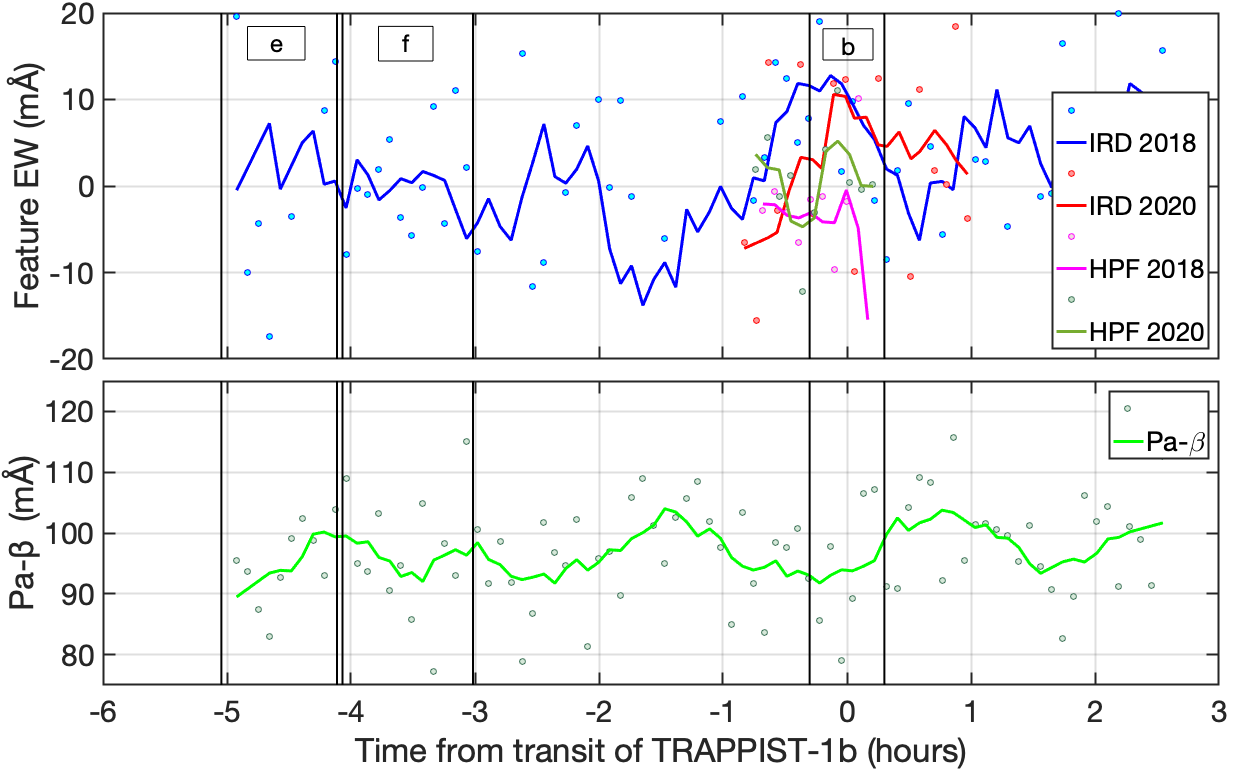}
\caption{{\bf Top}: Strength of the feature between 1083.285 and 1083.34\,nm (``\textit{region of interest}'') vs. time as measured in IRD (blue and red) and HPF (magenta and green) spectra.   The solid lines represent 7-point, first-order Savitzky-Golay filtered versions. {\bf Bottom}: EW of Hydrogen Paschen beta line as measured in IRD spectra in 2018. Both panels correlates negatively.}
\label{fig:ew_time}
\end{figure}

This feature is not due to telluric variations over transit/airmass as we do not have any known telluric \unit{H_2O} and OH line in the vicinity (Figure \ref{fig:fluxratio}). It could also have its origin from planetary occultation of active regions of the star responsible for the He\,I signal. However, we rule this out because we could reproduce the feature from two transits of planet `b' through IRD as seen in temporal variation plot (Figure \ref{fig:ew_time}: top panel).
Changes of the stellar line shape due to active regions appearing or disappearing from the stellar limb could also mimic this signal. But we consider this to be unlikely as the stellar He\,I line is not detected and the star's rotation period \citep[3.295 days;][]{2017ApJ...841..124V} is long compared to the transit duration. 

The observations of TRAPPIST-1e do show some variation in their in-transit spectra when compared to their out-of-transit spectra (see right panels of Figure \ref{fig:fluxratio}). By performing a similar random sampling analysis, we find that the mean-strength of the feature during the transit of TRAPPIST-1e (EW=5.41$\pm$2.28 m\AA{}) is higher than the mean-strength during the out-of-transit (EW=0.11$\pm$3.79 m\AA{}). However, such features are not seen during the transit of TRAPPIST-1f in the IRD spectra.

TRAPPIST-1 being a fainter target, causes excess scatter in the RV data \citep{2020ApJ...899L..13H}. Also TRAPPIST-1 is prone to have strong activity on its surface \citep{2017MNRAS.465L..74W, 2017ApJ...841..124V} and with lack of simultaneous measurement of activity indicators (e.g., Hydrogen Balmer lines) we could not rule not the feature's origin from stellar activity. Since the stellar He\,I lines are not detected, we chose a $\sim$0.55 \AA{} wide region as a ``\textit{region of interest}'' around the strongest doublet (1083.285-1083.34 nm in vacuum; Figure \ref{fig:fluxratio}) to compute upper limits. To limit any planet-related He\,I absorption in planet `b', `e' and `f', we modeled a simple Gaussian profile with a thermal broadening of 1000-30000 K. Running a MCMC simulation, we limit the Equivalent Width (EW) of any planet-associated absorption to 7.754 m\AA{}, 10.458 m\AA{}, and 4.143 m\AA{} for planets `b', `e', and `f', respectively, from the IRD data, and 3.467 m\AA{} for planet `b' from HPF data at 95\% confidence.

\subsubsection{Temporal variation}

The Paschen beta (Pa-$\beta$) emission line is considered as a tracer for outflow activity in T-Tauri stars \citep{2004A&A...417..247W}. In our observations, Pa-$\beta$ is seen as an absorption line. The temporal variation of the strength of ``\textit{region of interest}'' with the Pa-$\beta$ from the IRD 2018 spectra is shown in Figure \ref{fig:ew_time}. The EW of the feature in ``\textit{region of interest}'' from IRD does not correlate with the Pa-$\beta$ line. The insignificant negative correlation coefficient (Pearson's $R = -0.0438$ and $p = 0.6905$; Spearman Rank $\rho = -0.0171$ and $p = 0.8766$) could have arisen from the large scatter in the spectra, emphasizing the need for more observations. The 7-point, first-order Savitzky-Golay filtered version correlates with a better significance (Pearson's $R = -0.2403$ and $p = 0.0267$; Spearman Rank $\rho = -0.2275$ and $p = 0.0365$). The HPF data too shows a negative correlation with the Pa-$\beta$ feature (Pearson's $R = -0.4097$ and $p = 0.2108$; Spearman Rank $\rho = -0.4727$ and $p = 0.1456$). The Savitzky-Golay filtered version of HPF He\,I data does not correlate with the Pa-$\beta$ feature, largely due to the absence of post-egress data from the transit night.

\subsection{He I line modeling} \label{sec:hei_model}

We compared our upper limits on the transit-associated absorption in the 1083\,nm He\,I line to the predictions of a model of an escaping planetary atmosphere as an isothermal Parker wind which has previously been described \citep{2020MNRAS.495..650G, 2020MNRAS.498L.119G}.  The wind is assumed to have a solar composition (91.3\% H by number).  The metastable triplet (2$^3$S) state responsible for the \hei\ transition is, in these low density winds, populated by recombination of He ionized by EUV photons with energies $>26.4$ eV ($\lambda < 504$\AA) and depopulated by collisional-de-excitation.  For cool, inactive stars like TRAPPIST-1, ionization by near ultraviolet (NUV) photons with energies $>4.8$ eV \AA{} \citep[$\lambda < 2583$ \AA,][]{2018ApJ...855L..11O} is a minor triplet He sink.  Thus, the EUV spectral energy distribution of the host star is needed for such calculations.  That spectrum is not available for TRAPPIST-1 because of its intrinsic faintness, absorption by the ISM, and the lack of sufficiently sensitive space telescopes operating in the UV range.  Instead, we used model spectra of TRAPPIST-1 computed by \citet{Peacock2019} which includes emission from the photosphere, chromosphere, and transition region of the star, but not the corona.  \citet{Peacock2019} computed three models; one which matched a reconstruction of emission in the Lyman-$\alpha$ line based on \emph{HST} observations \citep{2017A&A...599L...3B}, and two that span the observed range of emission from nearby M8-type stars in the NUV channel of \emph{GALEX} (175-280\,nm), and upper limits in the far-ultraviolet (FUV, 135-175\,nm) channel.   The He\,I EWs calculated using these three models were nearly indistinguishable, and we show only those using the Ly-$\alpha$-calibrated model (Model 1). Planet and star parameters were taken from \citet{2020arXiv201001074A}. The two variable inputs to the wind model were the mass loss rate (ranging over 0.03-3 M$_{\oplus}$ Gyr$^{-1}$) and the wind temperature (ranging over 2000-20000\,K).  Estimates of current mass loss rates used both the maximal energy-limited rate (with 100\% efficiency) and model-derived rates from \citet{Kubyshkina2018}, combining estimates of the X-ray and EUV luminosities from \citet{Fleming2020} and the Lyman-alpha luminosity of \citet{2017A&A...599L...3B}.  Both sets of estimates are an order of magnitude higher than the mass-loss estimates of \citet{Becker2020}, who assume a fixed energy efficiency of 10\%.  

\begin{figure*}
\centering
\includegraphics[width=\textwidth]{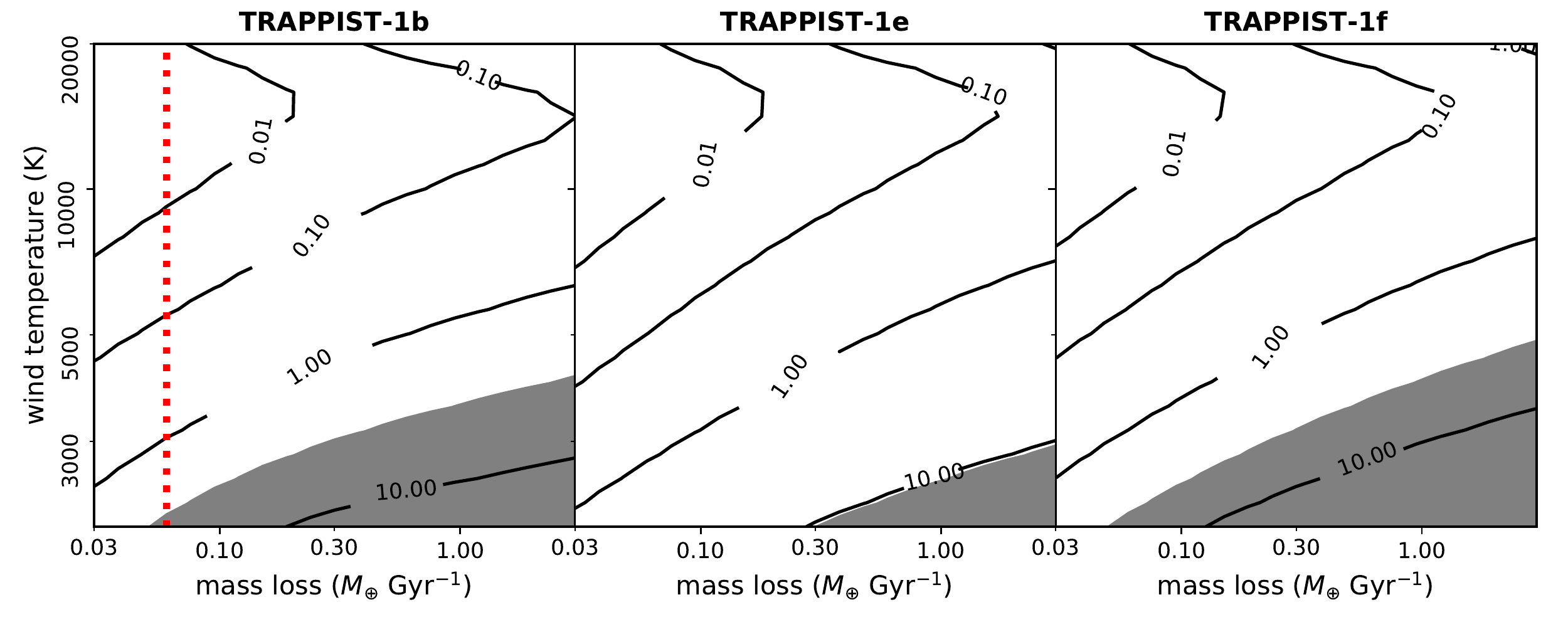}
\caption{Isothermal Parker wind model predictions of the strength (equivalent width) of the 1083\,nm He\,I line produced by hypothetical escaping atmosphere from TRAPPIST-1 planets ``b'', ``e'', and ``f'' (left to right), as a function of the mass loss rate and wind temperature.  The grey regions are the parameter spaces ruled out by our 95\% confidence upper limits on the EW of any transit-associated absorption. The vertical red dotted line in the plot of ``b'' is the energy-limited escape rate for 100\% efficiency. The predicted rates for ``e'' and ``f'' are too low to appear on the plots.}
\label{fig:hei_model}
\end{figure*}

Our upper limits on transit-associated He\,I absorption can only rule out the high mass-loss rate and low wind temperature corner of the parameter space (mass loss rates $\gtrsim  0.1 M_{\oplus}$ Gyr$^{-1}$ and $T_w \lesssim 4000$K (Fig. \ref{fig:hei_model}).  Except, marginally, for the case of planet ``b'', they do not constrain mass loss rate predictions.  For ``b'', our models exclude the highest possible (energy-limited) rate, and only for $T_w < 5000$K.  The low escape rates and corresponding He\,I EW is due to the low XUV luminosity of the small size and low activity of TRAPPIST-1, as well as the relatively high gravity of the Earth-size, Earth-mass planets compared to hydrogen-rich Neptune-like planets.  We also emphasize that our calculations of the 1083\,nm line assume a solar-like composition for the wind, which is not expected for Earth-size planets with masses and radii that are consistent with a rocky composition \citep{2020arXiv201001074A} and that orbit close to the host star.

\section{Discussion and Summary} \label{sec4}

Using the high resolution NIR spectra obtained of TRAPPIST-1b, e, and f with IRD/Subaru and HPF/HET, we attempt to place an upper limit of their He\,I 1083\,nm absorption during transits. Our analysis of the transit observations of TRAPPIST-1b, e and f did not detect the presence of low mean molecular-weight extended atmosphere in these planets. The estimated equivalent width (EW) upper limits are from the transit features seen in the ``\textit{region of interest}'', which may or may not be of planetary origin. In addition, the significance of the features in the ``\textit{region of interest}'' are within 2-$\sigma$ of noise level for planet `b' and `e' and within 1-$\sigma$ of noise level for planet `f'.

Our estimated upper limits for atmospheric escape rate translate to atmospheres extending up to 5.1, 13 and 2.4 planetary radii for planets TRAPPIST-1b, e and f respectively. Such extended atmospheres of planet `b' and `e' almost reach the planet's Roche radius whereas the limit of planet `f' is well within its Roche radius. 

\subsection{Stellar He I lines}

The stellar Helium absorption lines at 1083\,nm were not detected in both the IRD and HPF spectra. This can be attributed to the low temperature of the star. With large number of possible molecular lines in the vicinity, the true continuum level is difficult to estimate. With an approximated pseudo-continuum level, we do not see the lines in emission too as expected for an inactive star like TRAPPIST-1 \citep{2020arXiv201001074A}. This non-detection of stellar He\,I lines is evidence for a true, physical disappearance
of the line as opposed to an increasing level of fill-in caused, for
example, by collisional excitation due to raising activity levels \citep{2019A&A...632A..24F}. 

The features we see in the ``\textit{region of interest}'' could have originated from the stellar flares. But we do not see any trademarks of flares (sudden steep rise in flux and a long decay) during our full-night observation through IRD on UT 2018 August 30 (see bottom panel in Figure \ref{fig:ew_time}). Observation duration of both the transits from HPF and the 2020 IRD transit was small ($\sim$ 1-2 hours) and the Paschen-$\beta$ line was contaminated with telluric OH-emission line to get any useful information. Although TRAPPIST-1 is known to host high-energy flares (see \cite{2017ApJ...841..124V}), the lack of one during our observation diverts our attention towards other processes for its origin. We do acknowledge that there could be other stellar processes that could cause this variability, the analysis of which is beyond the scope of this study.

\subsection{Temporal analysis with Pa-$\beta$ line} \label{sec4sub1}

The EW of Pa-$\beta$ absorption line was computed for the 2018 IRD observation. The Pa-$\beta$ line from 2020 IRD observations were redundant as they were contaminated by a telluric OH-emission line. The HPF Pa-$\beta$ data was not usable as it lacked the pre-ingress and post-egress data, making it impossible to extract any useful information in the temporal analysis. 

There is no significant correlation of the feature in ``\textit{region of interest}'' with Pa-$\beta$ (see Figure \ref{fig:ew_time}). But, the filtered version of IRD spectra does show better negative correlation with Pa-$\beta$ than the raw data. However, we do emphasize the need for additional transit observations to get a more robust conclusion on the correlation between the possible He\,I excess absorption and Pa-$\beta$ as the physical significance of this correlation are not well understood.

\subsection{Helium-dominated atmospheres?}

The recent long-term observation of TRAPPIST-1 \citep{2020arXiv201001074A} showed that the planet TRAPPIST-1b, e and f have a significantly high density when compared to the previous estimates \citep[e.g.,][]{2018A&A...613A..68G}. Also the \textit{HST} observations presented in \cite{2016Natur.537...69D, 2018NatAs...2..214D} also indicate the absence of cloud-free hydrogen-dominated atmospheres in the three TRAPPIST-1 planets in our current study. However, the TRAPPIST-1 planets might currently have Helium-dominated upper atmospheres. After the initial XUV and EUV dominated hydrodynamic evaporation for $\sim$ $0.1-1$ Gyr, if TRAPPIST-1 planets were born with H/He atmospheres in the order of $10^{-3}$ planetary mass, their atmospheres might have depleted in Hydrogen but still be abundant in Helium even after $\sim$ 10 Gyr \citep{2015ApJ...807....8H, 2020ApJ...889...77H}. However, there might not be sufficient high-energy radiation available at the planet's orbital distance, even for the closest planet `b' to produce this 1083 nm He\,I feature.

Using the equation 1 from \cite{2017MNRAS.465L..74W}, with heating-efficiency as $\mathrm{\eta = 100\%}$ and reduced energy to escape the Roche Lobe as $\mathrm{K_{tide} = 1}$ \citep{2007A&A...472..329E}, for an energy-limited escape, the flux needs to be at least 2 orders of magnitude higher than the estimate from \cite{Peacock2019} for the closest planet `b'. The lack of visible difference during the transit of TRAPPIST-1b between in-transit and out-of-transit spectrum in the ``\textit{region of interest}'' might be indicating the absence of detectable-level evaporating Helium in the atmosphere.

\subsection{Atmosphere escape modeling}

Our model assumes a solar-like composition for the Parker wind, which might not be applicable to the close-in TRAPPIST-1 planets. Nevertheless, the model still uses the best available estimate of EUV flux of TRAPPIST-1 \citep{Fleming2020, 2017MNRAS.465L..74W, 2017A&A...599L...3B}. The model also assumes the wind to be of solar composition (91.3\% H by number), which might be drastically different from the higher Helium concentrated atmosphere. The higher Helium concentration would increase the mean molecular weight, thereby lowering the wind mass loss rate. The updated SED of TRAPPIST-1 presented recently in \cite{2021ApJ...911...18W} shows a lower XUV and EUV flux compared to the Lyman-$\alpha$ scaled model from \cite{Peacock2019}. This translates to higher mass-loss rates and even more weaker mass-loss constrains than the results presented in this paper.

\subsection{Summary}

We have presented the non-detection upper limit on the equivalent widths of <7.754 m\AA{} for planet `b', <10.458 m\AA{} for planet `e', and <4.143 m\AA{} for planet `f' at 95\% confidence from the IRD data, and <3.467 m\AA{} for planet `b' at 95\%i confidence from HPF data. The complete EUV spectral energy distribution is not available for any ultra-cool star \citep{Peacock2019}. Also, the Helium 2$^3$S state are expected to be not populated efficiently for late M-dwarf like TRAPPIST-1 \citep{2019ApJ...881..133O}, making them a poor target for Helium observations. However, the close proximity of the planets to the host star makes the TRAPPIST-1 system an unique target among the cool stars to ``observationally'' constrain the possible atmospheric escape. Nevertheless, we emphasize that more observations are needed to obtain a robust constrain on the presence or absence of Helium abundant atmospheres. 

Our estimates on atmospheric escape, although not constrained, will aid in further characterization of high mean molecular weight lower atmospheres, if present, in TRAPPIST-1 planets. The upcoming James Webb Space Telescope's mid-infrared instrument NIRSpec is expected to probe such species with spectral signatures at infrared wavelengths.

\acknowledgements
The \citet{Peacock2019} model UV spectra were obtained from the MUSCLES Treasury survey on the Mikulski Archive for Space Telescopes. V.K would like to thank Japanese Government's MEXT Scholarship.

This work was partially supported by JSPS KAKENHI Grant Number JP19K14783. M.T. is supported by JSPS KAKENHI grant Nos. 18H05442, 15H02063, and 22000005. Y.H. is supported by JSPS KAKENHI grant No. 18H05439. CIC is supported by NASA Headquarters under the NASA Earth and Space Science Fellowship Program through grant 80NSSC18K1114. R.K acknowledges support from the GSFC Sellers Exoplanet Environments Collaboration (SEEC), which is supported by NASA’s Planetary Science Division’s Research Program. EG is supported by the NASA Exoplanets Research Program (Award 80NSSC20K0957). 

This work was partially supported by funding from the Center for Exoplanets and Habitable Worlds. The Center for Exoplanets and Habitable Worlds is supported by the Pennsylvania State University, the Eberly College of Science, and the Pennsylvania Space Grant Consortium. This work was supported by NASA Headquarters under the NASA Earth and Space Science Fellowship Program through grants 80NSSC18K1114. We acknowledge support from NSF grants AST-1006676, AST-1126413, AST-1310885, AST-1517592, AST-1310875, AST-1910954, AST-1907622, AST-1909506, the NASA Astrobiology Institute (NAI; NNA09DA76A), and PSARC in our pursuit of precision radial velocities in the NIR. Computations for this research were performed on the Pennsylvania State University’s Institute for Computational \& Data Sciences (ICDS). This work was supported by NASA Headquarters under the NASA Earth and Space Science Fellowship Program through grants 80NSSC18K1114. These results are based on observations obtained with the Habitable-zone Planet Finder Spectrograph on the Hobby-Eberly Telescope. We thank the Resident astronomers and Telescope Operators at the HET for the skillful execution of our observations of our observations with HPF. The HET is a joint project of the University of Texas at Austin, the Pennsylvania State University, Ludwig-Maximilians-Universität München, and Georg-August Universität Gottingen. The HET is named in honor of its principal benefactors, William P. Hobby and Robert E. Eberly. The HET collaboration acknowledges the support and resources from the Texas Advanced Computing Center. Authors would also like to thank the anonymous reviewer for his/her constructive comments. 
\facilities{IRD/Subaru, HPF/HET, ARCTIC/APO} 

\software{AstroImageJ \citep{collins2017},
\texttt{astropy} \citep{astropy2013},
\texttt{astroscrappy} \citep{astroscrappy},
\texttt{batman} \citep{kreidberg2015},
\texttt{barycorrpy} \citep{kanodia2018}, 
\texttt{celerite} \citep{Foreman-Mackey2017},
\texttt{GNU Parallel} \citep{Tange2011}, 
\texttt{HxRGproc} \citep{ninan2018},
\texttt{juliet} \citep{Espinoza2019}.}

\bibliography{trappist_1b_bib,gaidos_refs.bib,hpf_refs.bib}{}
\bibliographystyle{aasjournal}

\end{document}